\documentstyle[12pt,psfig]{article}
\let\section=\subsection     \let\subsection=\subsubsection                

\begin{document}
\begin{center}
   {\large \bf SIMULATION OF HEAVY ION REACTIONS WITH}\\[2mm]
   {\large \bf NONLOCAL KINETIC EQUATIONS}\\[5mm]
   K. Morawetz$^*$, V. \v Spi\v cka$^{**}$, P. Lipavsk\'y$^{**}$, Ch. Kuhrts$^*$\\[5mm]
   {\small \it  $^*$  Fachbereich Physik, University Rostock, D-18051 Rostock, Germany \\
\small \it 
$^{**}$ Institute of Physics, Academy of Sciences, Cukrovarnick\'a 10,
16200 Praha 6, Czech Republic\\[8mm] 
}
\end{center}

\begin{abstract}\noindent
The incorporation of realistic trajectories of two-particle scattering into existing BUU or QMD codes meets considerable difficulties. We propose a method of replacing the scattering event by a tractable non-local scenario reproducing the correct asymptotics. The first principle derivation justifies the use of corresponding nonlocal kinetic equations. The nonlocal shifts are necessary if the system should approach an equation of state with second quantum virial coefficient. The modifications of QMD and BUU codes are discussed and comparison are presented with resent data. We find a pre-equilibrium production of particles leading to higher energetic distributions of protons.
\end{abstract}

\section{Introduction}
Recent dynamical simulations of heavy ion reactions at low and mid energies are based either on BUU or QMD simulations. 
These equations simulate the Boltzmann equation (BE) with the extension of Pauli-blocking and meanfield drift
\begin{eqnarray}
&&{\partial f_1\over\partial t}+{\partial\varepsilon_1\over\partial k}
{\partial f_1\over\partial r}-{\partial\varepsilon_1\over\partial r}
{\partial f_1\over\partial k}
=\sum_b\int{dpdq\over(2\pi\hbar)^3} {1\over \mu^2} {d \sigma \over d \Omega}
\delta\left(\varepsilon_1+\varepsilon_2-
\varepsilon_3-\varepsilon_4\right)
\nonumber\\
&&\times      
\Bigl[f_3f_4\bigl(1-f_1\bigr)\bigl(1-f_2\bigr)-
\bigl(1-f_3\bigr)\bigl(1-f_4\bigr)f_1f_2\Bigr].
\label{1}
\end{eqnarray}
The arguments of distributions $f$ and energies $\varepsilon$ are shortened
as $f_1\equiv f_a(k,r,t)$, $f_2\equiv f_b(p,r,t)$,
$f_3\equiv f_a(k-q,r,t)$, and $f_4\equiv f_b(p+q,r,t)$, with momenta
$k,p,q$, coordinate $r$, time $t$, spin and isospin $a,b$ and reduced mass $\mu$. 

While the motion due to nonlinear meanfield is governed by the quasiclassical Hamilton equations for chosen test particles or wave packets, the collision is performed randomly if the particles meet inside the area of cross section or at the closest approach. By random choice of transfer momenta and averaging over many runs the scattering probability given by the cross section in the Boltzmann equation is reproduced. Medium effects are discussed in terms of density and temperature dependent depletion of cross section.

However, there are modifications of the Boltzmann equation which have a more fundamental meaning.
It has been noticed that the starting Boltzmann equation has an inherent contradiction. While the collision probability describes a finite diameter the collision takes place at a time instant and local in space. This deficiency has to be cured. Simple cases like hard sphere gases have been discussed already be Enskog \cite{CC90} and in nuclear matter by Malfliet \cite{M83} and Halpert \cite{H81}. Resonance scattering has been discussed by Pratt and Danielewicz \cite{DP96}.
The aim of these corrections are to describe a realistic scattering event within kinetic theory in a way that the correct asymptotics are reproduced. Only by this way the virial corrections, e.g. the two-particle correlation energy in balance equations can be expected to be approached during time evolution. The original Boltzmann equation and consequently the standard BUU and QMD simulations lead to no binary correlation energy. As it was shown \cite{SLM96} this binary correlation can be reproduced within a non-local collision integral.

The need for nonlocal corrections can be stimulated by discussing the scattering of two particles as superpositions of wave packets \cite{SLMa96} and similar \cite{AH97}. The solution of the free Schroedinger equation can be written for the relative motion with c.m. momentum $k$ as
\begin{eqnarray}
\phi(p,k,t)=F(p,k) {\rm e}^{-i E_p t}
\end{eqnarray}
and the asymptotic wave after scattering can be decomposed into $\Psi=\phi+\phi^{\rm sc}$ with the scattered wave for large distance $x$ from scattering center
\begin{eqnarray}
\phi^{\rm sc}(x,k,t)&=&\int {d p\over (2 \pi \hbar)^3} {\cal F} (p,k) {f(p,\cos(px))\over x} {\rm e}^{i (p x -E_p t)}
\label{p1}
\end{eqnarray}
with the scattering amplitude $f(p,\cos (p x))$. We proceed now and expand the scattering amplitude around the center of mass momentum $k$
\begin{eqnarray}
&&f(p,\cos(px))=|f(p,\cos (px))|{\rm e}^{i\delta(p,\cos(px))}\nonumber\\
&&=f(k,\cos(kx)) \left (1+ (p-k) \nabla_p |f(p,\cos(px))|_{p=k}\right ) {\rm e}^{i (p-k) \nabla_p \delta(p,\cos(px))_{p=k}}.\nonumber\\
&&
\end{eqnarray}
The derivative of the phase $\delta$ leads now to the definition of the effective space shifts $\Delta$ and the time shift $\Delta_{||}$
\begin{eqnarray}
2\Delta=\nabla_p\delta=\partial_k \delta+{{x\over |x|}-\cos(xk){k\over |k|}
\over |k|}\partial_{\cos(xk)}\delta\equiv {k\over m}\Delta_{\|}+\Delta_{\perp}
\label{pp}
\end{eqnarray}
where we denoted the shifts corresponding to the direction of $k$ as $\|$ and $\perp$. Expanding in (\ref{p1}) also $E_p=p^2/m=k^2/m +(p-k)2 k/m$ and collecting all terms we obtain
\begin{eqnarray}
\phi^{\rm sc}(x,k,t)={f(k,\cos(kx))\over x}\int {dp\over (2 \pi \hbar)^3} \tilde 
{\cal F}(p,k) {\rm e}^{ip(x+2 \Delta-u t)}{\rm e}^{-i {k^2\over m} (\Delta_{\|}-t)}
\end{eqnarray}
where the wave packet velocity is $u=2 k/m$. We observe three effects of scattering on the asymptotics: (i) a genuine time delay $\Delta_{\|}$, (ii) an effective displacement of the two colliding particle of $\Delta$ with respect to the center of mass and (iii) a modification of scattering probability 
$\tilde{\cal F}={\cal F} (1+ (p-k) \nabla_p |f(p,\cos(px))|_{p=k})$. 

In this contribution we will put these ideas of nonlocalities
on the firm ground using the quantum kinetic equation with
nonlocal scattering integrals which was derived from quantum
statistics \cite{SLM96,LSM97} to show how the effect of nonlocalities play
a role in simulations of heavy ion reactions and compare them with experiment.

\section{Nonlocal kinetic equation}
In \cite{SLMa96,SLM96,LSM97} we have given a systematic derivation of nonlocal kinetic equation (NKE) which includes hard sphere like corrections, resonance like corrections as well as rotational angle corrections, which turned out to be important for angular momentum conservation. This kinetic equation completes all balance equatons on the level of quantum second virial coefficients. 
The NKE-equation with the
collected gradient terms reads
[$\Delta_r={1\over 4}(\Delta_2+\Delta_3+\Delta_4)$]
\begin{eqnarray}
&&{\partial f_1\over\partial t}+{\partial\varepsilon_1\over\partial k}
{\partial f_1\over\partial r}-{\partial\varepsilon_1\over\partial r}
{\partial f_1\over\partial k}
=\sum_b\int{dpdq\over(2\pi)^5}\delta\left(\varepsilon_1+\varepsilon_2-
\varepsilon_3-\varepsilon_4+2\Delta_E\right)
\nonumber\\
&&\times 
z_1z_2z_3z_4
\Biggl(1-{1\over 2}{\partial\Delta_2\over\partial r}
-{\partial\bar\varepsilon_2\over\partial r}
{\partial\Delta_2\over\partial\omega}\Biggr)
\nonumber\\
&&\times
|T_{\rm sc}^R|^2\!\left(\varepsilon_1\!+\!\varepsilon_2\!-\!
\Delta_E,k\!-\!{\Delta_K\over 2},p\!-\!{\Delta_K\over 2},
q,r\!-\!\Delta_r,t\!-\!{\Delta_t\over 2}\!\right)
\nonumber\\
&&\times\Bigl[f_3f_4\bigl(1-f_1\bigr)\bigl(1-f_2\bigr)-
\bigl(1-f_3\bigr)\bigl(1-f_4\bigr)f_1f_2\Bigr]\nonumber\\&&
\label{9}
\end{eqnarray}
with the nonlocal shifts given as derivatives of the total phase shift
of the scattering $T$-matrix 
$\phi={\rm Im\ ln}T^R_{\rm sc}(\Omega,k,p,q,t,r)$,
\begin{equation}
\begin{array}{lclrclrcl}
\Delta_2&=&
{\displaystyle\left({\partial\phi\over\partial p}-
{\partial\phi\over\partial q}-{\partial\phi\over\partial k}
\right)}&\ \ 
\Delta_3&=&
{\displaystyle\left.-{\partial\phi\over\partial k}
\right .}&\ \ 
\Delta_4&=&
{\displaystyle-\left({\partial\phi\over\partial k}+
{\partial\phi\over\partial q}\right)}
\\ &&&&&&&&\\ 
\Delta_t&=&
{\displaystyle \left.{\partial\phi\over\partial\Omega}
\right .}&\ \
\Delta_E&=&
{\displaystyle \left.-{1\over 2}{\partial\phi\over\partial t}
\right .}&\ \ 
\Delta_K&=&
{\displaystyle \left.{1\over 2}{\partial\phi\over\partial r}
\right .}.
\end{array}
\label{8}
\end{equation}
After derivatives, $\Delta$'s are evaluated at the energy shell
$\Omega\to\varepsilon_3+\varepsilon_4$.
Unlike in (\ref{1}), the subscripts denote shifted arguments:
$f_1\equiv f_a(k,r,t)$, $f_2\equiv f_b(p,r\!-\!\Delta_2,t)$,
$f_3\equiv f_a(k\!-\!q\!-\!\Delta_K,r\!-\!\Delta_3,t\!-\!\Delta_t)$, and
$f_4\equiv f_b(p\!+\!q\!-\!\Delta_K,r\!-\!\Delta_4,t\!-\!\Delta_t)$.

\begin{figure}
\parbox[h]{14.5cm}{
\parbox[]{8cm}{
  \psfig{figure=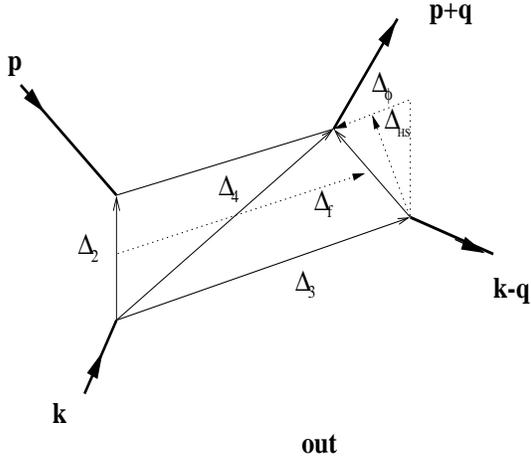,width=7cm,height=6cm}}
\hspace{1cm}
\parbox[]{5cm}{\vspace{2cm}\caption{The equivalent scattering event of two colliding particles.\label{soft}}}
}
\end{figure}

The $\Delta$'s are effective shifts and they represent mean values of
various nonlocalities of the scattering integral as demonstrated in figure \ref{soft}. These shifts enter
the scattering integral in the form known from the theory of gases
\cite{CC90,NTL91,H90}, however, the set of shifts is larger due to the
medium effects on the binary collision that are dominated by the Pauli
blocking of the internal states of the collision. 

Now we want to use these derived nonlocalities to mimick the
real scattering event in simulations. To this end we use
a classical analogy of these nonlocalities and the scattering
event as demonstrated by picture \ref{soft}. We obtain the following 
scenario: Two particles approach until they reach a distance $\Delta_2$. Then they form a molecule traveling over a distance $\Delta_f$ with a time $\Delta_t$. During this propagation the molecule rotate about $\Delta_\phi$.
Now the kinetic equation (\ref{9}) is numerically tractable by recent Monte
Carlo or BUU codes \cite{KDB96} as demonstrated below.

\section{Instant approximation}

The selfconsistent evaluation of all
$\Delta$'s for all collisions would be too demanding. We employ two
kinds of additional approximations. First, following approximations used
within the BUU equation, we neglect the medium effect on binary
collision, i.e., use the well known free-space T-matrix. Second, we
rearrange the scattering integral into an instant but non-local form.
This instant form parallels hard-sphere-like collisions what allow us
to employ computational methods developed within the theory of gases
\cite{AGA95} similarly as it has been done in \cite{KDB96}.

\begin{figure}
\parbox[h]{14.5cm}{
\parbox[]{8cm}{
  \psfig{figure=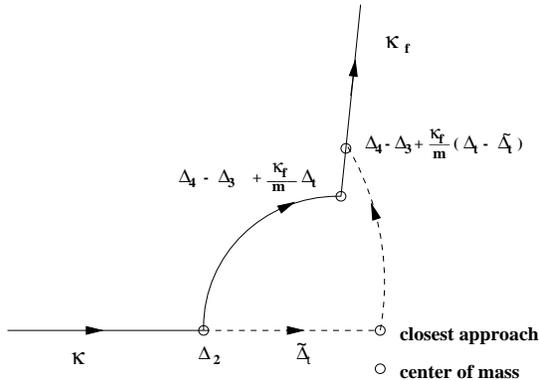,width=7cm,height=5cm}}
\hspace{1cm}
\parbox[]{5cm}{\vspace{3cm}\caption{A nonlocal binary collision (solid line) together with the scenario of sudden jump at the closest approach.
\label{softc}}
}}
\end{figure}

In the instant approximation we let particles to make a sudden jump at
time $t$ from $r_a$ and $r_b$ to effective final coordinates $\tilde
r_a$ and $\tilde r_b$. These effective coordinates and momenta
$\tilde\kappa$ and $\tilde K$ are selected so that at time $t+\Delta_t$
particles arrive at the correct coordinates, $r_a'$ and $r_b'$, with the
correct momenta, $\kappa'$ and $K'$. Accordingly, in the asymptotic
region, after $t+\Delta_t$, there is no distinction between the
non-instant and instant pictures, which is shown as solid line in figure \ref{softc}. 
This asymptotic condition is naturally
met if one extrapolates the out-going trajectories from known
coordinates and momenta at $t+\Delta_t$ back to the time $t$. Doing so
one finds that the effective coordinates read
\begin{eqnarray}
\tilde r_a&=&r_a'-{k-q\over m}\Delta_t=
r_a+\Delta_3-{k-q\over m}\Delta_t,
\label{da}\\
\tilde r_b&=&r_b'-{p+q\over m}\Delta_t=
r_b+\Delta_4-\Delta_2-{p+q\over m}\Delta_t.
\label{db}
\end{eqnarray}

When incorporating the displacements into the QMD simulation code, we
have to face the fact that two particles are selected for a collision if they meet at the point of closest approach.
This distance is different from the distance $\Delta_2$ required from the equivalent scattering scenario presented in figure \ref{softc} as solid line. We consider now the time required to travel from $\Delta_2$ to the distance of closest approach $\tilde\Delta_t={m\over 2 \kappa^2} \kappa \Delta_2$ in analogy to \cite{T81}. Within this scenario we are allowed to jump at the point of closest approach to the final asymptotics (\ref{da}) and (\ref{db})with the additional distance the particle travel during $\tilde\Delta_t$. The effective final coordinates thus have to be evaluated as
\begin{equation}
\tilde r_{a,b}={R_a+R_b \over 2}\mp\Delta,
\label{df}
\end{equation}
with the effective displacement
\begin{equation}
\Delta=\frac 1 2 \Delta_2-\Delta_3+{k-q\over m}(\Delta_t-\tilde\Delta_t).
\label{dd}
\end{equation}
Since the center of mass does not jump in the collision, the final
displacement can be also written in an alternative way, $\Delta
=\frac 1 2 \Delta_2+\Delta_4-\Delta_2-{p+q\over m}(\Delta_t-\tilde\Delta_t)$. The
non-local corrections are thus performed as follows. When the collision
is selected, we evaluate $\Delta$ from (\ref{dd}) and
(\ref{8}), redisplay particles into $\tilde r_a$ and $\tilde r_b$
and continue with the simulation.

Let us note that from (\ref{8}) and (\ref{dd}) the already presented shift (\ref{pp}) follows as it should.

At this point it is possible to establish a connection of the present
theory to the hard-sphere-like corrections used by Malfliet \cite{M83}
and Kortemeyer, Daffin and Bauer \cite{KDB96}. For hard spheres of the
diameter $d$, the phase shift has a classical limit $\phi=\pi-|q|d$
which gives $\Delta_3=0$ and $\Delta_2=\Delta_4=
{q\over|q|}d$. The displacement thus has the same amplitude $d$ for all
binary collisions and points in the direction of the transferred
momentum, as it is known from the Enskog equation \cite{CC90}.

\section{Results for Simulation}
In order to investigate the effect of non-local shifts on realistic
simulations of a heavy ion reaction, we have evaluated $\Delta^{\rm f}$
from the two-particle scattering T-matrix $T^{R}$ in the Bethe-Goldstone 
approximation \cite{D84,SLM96} using the separable Paris potential \cite{HP84}.
The comparison of the shifts calculated for different potentials concerning partial wave coupling up to D-waves can be found in \cite{MLSK98}. 
We have incorporated these shifts into
a QMD code for the central collision of
$^{129}$Xe$\rightarrow$$^{119}$Sn at $50$~MeV/A. 

\begin{figure}
\parbox[t]{20cm}{\psfig{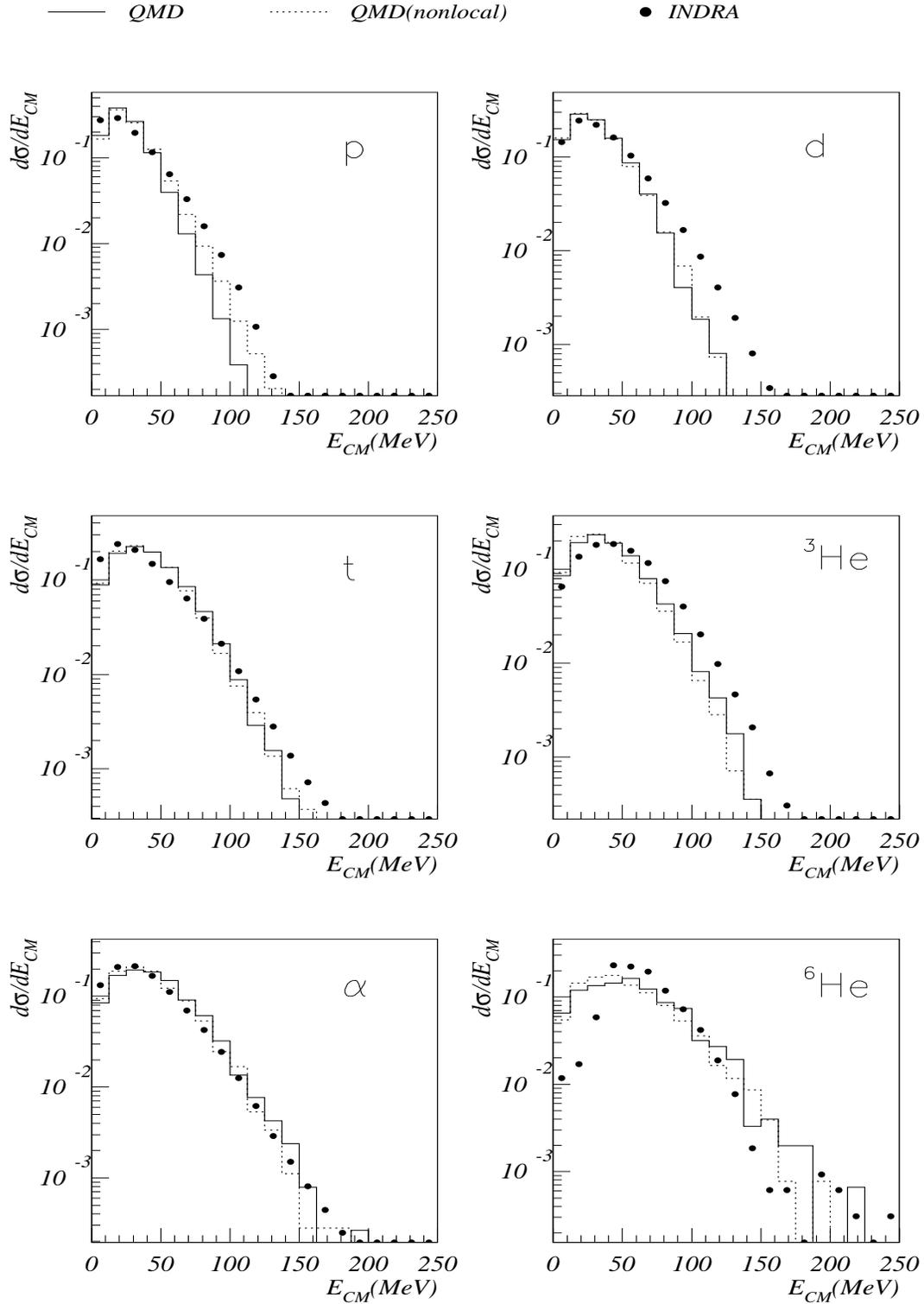}}
\caption{The particle spectra for central collision of
$^{129}$Xe$\rightarrow$$^{119}$Sn at $50$~MeV/A with and without
non-local corrections. The data are extracted from recent INDRA
experiments \protect\cite{INDRA}. The non-local corrections bring the spectrum
of the protons towards the experimental values leaving the
clusters almost unchanged.
\label{spek3}}
\end{figure}

Figure~\ref{spek3}a
shows the exclusive proton spectra subtracting the protons bound in
clusters. This procedure is performed within a spanning tree model which
is known to describe a production of light charged cluster in a
reasonable agreement with the experimental data, Figs.~\ref{spek3}b-f.
Within the local approximation, however, the remaining distribution of
high-energy protons is too low to meet the experimental values. As one
can see, the inclusion of non-local collisions corrects this shortage of
the QMD simulation. As demonstrated in Fig.~\ref{spek3}, productions of
light clusters are rather insensitive to the non-local corrections. This
also shows that the improvement of the proton production is not on cost
of worse results in other spectra.

\begin{figure}
\parbox[h]{15.5cm}{
\parbox[]{8cm}{
\psfig{file=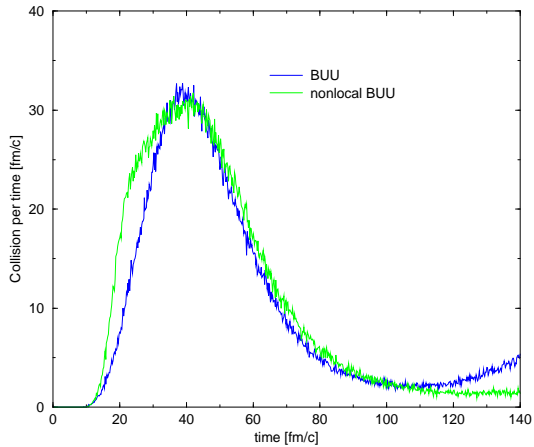,width=7cm,height=6cm,angle=-90}}
\hspace{1cm}
\parbox[]{5cm}{\vspace{3cm}\caption{The number of collisions per time with and without non-local
collisions within a BUU simulation of the same reaction as in figure
(\protect\ref{spek3}).
\label{n}}
}}
\end{figure}

A microscopic mechanism leading to the increase in the high-energy part
of the particle spectrum can be traced down to an enhancement of the
number of collisions at the pre-equilibrium stage of the heavy ion
reaction demonstrated in Fig.~\ref{n} for the BUU simulation of the same
reaction. This enhancement gives rise to an immediate proton production
which itself translates into a high energetic spectra. In other words,
the strong production of the high-energy protons follows from the
pre-equilibrium emission of particles. The BUU simulation also shows
that non-local corrections are important namely in the early stage of
reaction well before most of light clusters are formed. It explains why
the production of protons is affected while the formation of light
clusters is nearly untouched by the non-local corrections.

\section{Conclusion}

In summary, as documented by the improvement of the high-energy proton
production, the non-local treatment of the binary collisions brings a
desirable contribution to the dynamics of heavy ion reactions. According
to an experience from the theory of gases, one can also expect a vital
role of non-localities in the search for the equation of state of the
nuclear matter. It is encouraging that the non-local corrections are
easily incorporated into the BUU and QMD simulation codes and do not
increase computational time. Corresponding programs can be obtained from authors.

\medskip\noindent
We thank the INDRA collaboration for the use of data prior to
publication. J.Aichelin and W. Bauer are thanked for the well documented
QMD and BUU codes. This work was supported from Czech Republic, the GACR
Nos.~202960098 and 202960021, and GAAS Nr. A1010806, and Germany, the
BMBF, Nr. 06R0884, and the Max-Planck-Society.

\end{document}